\documentclass[useAMS,usenatbib,usegraphicx]{mn2e}
\usepackage{txfonts}
\DeclareMathAlphabet{\mathpzc}{OT1}{pzc}{m}{it}
\usepackage{epsfig}
\usepackage{color}
\font\sc=cmcsc10
\input epsf

\title[  Dwarf spheroidal galaxy kinematics and  spiral galaxy scaling laws]
       { Dwarf spheroidal galaxy kinematics and spiral galaxy scaling laws}
\author[P. Salucci et al.]
{Paolo Salucci$^1$,
Mark I. Wilkinson$^2$,
Matthew G. Walker$^{3,4}$, 
Gerard F. Gilmore$^5$,\newauthor
Eva K. Grebel$^6$,
Andreas Koch$^7$,
Christiane Frigerio Martins$^8$,
Rosemary F.G. Wyse$^9$\\\\
$^1$SISSA, Department of Astrophysics, via Beirut, 2-4, 34014  
Trieste, Italy\\
$^2$Department of Physics and Astronomy, University of Leicester,  
University Road, Leicester LE1 7RH, UK\\
$^3$ Harvard-Smithsonian Center for Astrophysics, 60 Garden St., Cam-
bridge, MA 02138\\
$^4$ Hubble Fellow\\
$^5$Institute of Astronomy, University of Cambridge, Madingley Road,  
Cambridge, CB3 OHA, UK\\
$^6$ Astronomisches Rechen-Institut, Zentrum f\"ur Astronomie der 
Universit\"at Heidelberg, M\"onchhofstr. 12-14, 69120 Heidelberg, Germany\\
$^7$Zentrum f\"ur Astronomie der Universit\"at Heidelberg, Landessternwarte,
K\"onigstuhl 12, 69117 Heidelberg, Germany\\
$^8$INFES, Universidade Federal Fluminense, av. Jo\~{a}o Jazbik. 28470 
Santo Ant\^{o}nio de P\'{a}dua, Rio de Janeiro, Brazil\\
$^9$ Johns Hopkins University, Baltimore, MD, USA
}

\begin{document}

\maketitle
\begin{abstract}
  Kinematic surveys of the dwarf spheroidal (dSph) satellites of the
  Milky Way are revealing tantalising hints about the structure of
  dark matter (DM) haloes at the low-mass end of the galaxy luminosity
  function. At the bright end, modelling of spiral galaxies has shown
  that their rotation curves are consistent with the hypothesis of a
  Universal Rotation Curve whose shape is supported by a cored dark
  matter halo. In this paper, we investigate whether the internal
  kinematics of the Milky Way dSphs are consistent with the particular
  cored DM distributions which reproduce the properties of spiral
  galaxies.  Although the DM densities in dSphs are typically almost
  two orders of magnitude higher than those found in (larger) disk
  systems, we find consistency between dSph kinematics and Burkert DM
  haloes whose core radii $r_0$ and central densities $\rho_0$ lie on
  the extrapolation of the scaling law seen in spiral galaxies:
  $\log\: \rho_0 \simeq \alpha \:\log\: r_0$ + const with $0.9
  <\alpha<1.1$. We similarly find that the dSph data are consistent
  with the relation between $\rho_0$ and baryon scale length seen in
  spiral galaxies. While the origin of these scaling relations is
  unclear, the finding that a single DM halo profile is consistent
  with kinematic data in galaxies of widely varying size, luminosity
  and Hubble Type is important for our understanding of observed
  galaxies and must be accounted for in models of galaxy formation.
\end{abstract}

\begin{keywords}
dark matter---galaxies: dwarf spheroidal---galaxies:
kinematics and dynamics---Local Group---stellar dynamics
\end{keywords}

\section{Introduction}
\label{sec:intro}
In the current cosmological paradigm, dark matter (DM) provides the
gravitational potential wells in which galaxies form and evolve.  Over
the past decades, observations have provided detailed information
about the distribution of DM within those regions of spiral galaxies
where the baryons reside
(\citeauthor{Ashman92}~\citeyear{Ashman92};~\citeauthor{PSS96}~\citeyear{PSS96}
(hereafter
PSS);~\citeauthor{sofue}~\citeyear{sofue};~\citeauthor{Salucci07}~\citeyear{Salucci07}).
Similar information on the distribution of DM is also available for
low-surface-brightness (LSB) galaxies~\citep{deblok05,K06}. In these
disk systems, the ordered rotational motions and known geometry of the
tracers has facilitated the mass modelling and provided clear evidence
that the stellar components of spiral galaxies are embedded in
extended DM haloes. In the most luminous objects the stellar disk is
almost self-gravitating with dark matter contributing significantly to
the dynamics only at larger radii. In contrast, at the faint end of
the galaxy luminosity function, baryons contribute a negligible amount
to the overall gravitational potential~\citep{deblok08}.

Extensive modelling of both individual and co-added spiral galaxy
rotation curves (RCs) has generally concluded that almost maximal
stellar disks embedded in cored dark matter haloes reproduce the data
better than models with cosmologically-motivated, cusped dark matter
haloes~\citep[PSS; ][see
  also~\citeauthor{chemin11}~\citeyear{chemin11}]{PSS96,sb01,gentile04,deblok01,deblok02,marchesini02,g05,g07}. Further,
scaling relations between properties of the spiral galaxies such as
central surface density, stellar scale radius and stellar velocity
dispersion have been identified and interpreted as signatures of the
physical processes which drive galaxy
formation~\citep[e.g.][]{Kormendy85,Kormendy87,Kormendy90,Burkert95,KF04}.

\cite{PS88} and \cite{PS91} demonstrated that the Burkert halo density profile
given by
\begin{equation}
\rho (r)={\rho_0\, r_0^3 \over (r+r_0)\,(r^2+r_0^2)}~,
\end{equation}
with two free parameters, the core radius $r_0$ and the central halo
density $\rho_0$, is consistent with the available kinematic data in
spiral galaxies. When the mass distribution in these galaxies is
modelled using the combined contributions of a~\cite{Freeman1970} disk
for the luminous matter and a Burkert profile for the dark matter
halo, the structural parameters obtained (DM central densities, core
radii, disk masses and length scales) exhibit a series of scaling
laws. This led to the hypothesis of a ``Universal Rotation Curve", an
empirical function of radius and luminosity that reproduces the RCs of
spiral galaxies~\citep[PSS; ][and references
therein]{PS88,PS91,Salucci07}.

In contrast, our knowledge of the mass distribution in
pressure-supported systems like elliptical galaxies is still limited
\cite[see][for a recent summary of the state of art]{Napolitano10}.
On-going observations of Local Group dwarf spheroidal galaxies (dSphs),
which occupy the faint end of the luminosity function of
pressure-supported systems, are currently yielding crucial information
about the properties of the dark and luminous components in these
objects and, in turn, on the underlying physical properties of DM
haloes~\protect\cite[e.g.][]{Tolstoy04,Gilmore07,strigari08,pena08,walker09a}.
It is, however, an intrinsically difficult task both observationally,
in terms of measuring velocities for sufficient numbers of tracers, as
well as from a dynamical modelling point of view, due the lack of
precise information on the dynamical state of the stellar populations.

The dSphs are typically at least two orders of magnitude less luminous
than the faintest spirals, and show evidence of being DM dominated at
all radii~\citep[e.g.][]{kleyna02}. Their typical stellar masses lie
in the range $3\times 10^5M_\odot$ to $2\times 10^7M_\odot$, although
the luminous masses of some recently discovered objects are as low as
$10^3M_\odot$~\citep{Martin2008}, while their stellar length scales
are of order $0.3$ kpc. In these systems, the DM halo typically
outweighs the baryonic matter by a large factor (from a few tens, up
to several hundred). An understanding of these objects is therefore
essential for understanding the nature of dark matter itself and to
build an observational picture of the outcome of galaxy formation on
small scales. Additionally, high-redshift dSph pre-cursors most likely
contributed significantly to the build-up of the stellar halo of the
Milky Way~\citep{helmi08}.  Given that the observed dSphs are
predominantly old, pressure-supported, spheroidal systems, their
evolutionary histories would be expected to differ significantly from
those of spirals, especially in the baryonic components~\citep[see
e.g.][]{Grebel2003}.

As in the case of spiral galaxies, a number of authors have found
evidence of universality in the DM halo properties of dSph galaxies.
\cite{Mateo98} found that the variation of the mass-to-light ratios of
dSphs with total luminosity was consistent with all dSphs containing
similar masses of dark matter within the volume occupied by their
stellar distributions.  This implies a larger proportion of dark
matter in the less luminous objects, a general characteristic of
spiral galaxy haloes~\citep{PS88,deblok08}. More recent analyses
\citep{Gilmore07,Koch07a,strigari08,walker09b}, based on extended
velocity dispersion profiles rather than central velocity dispersions
alone, have generally supported this conclusion.

A number of important questions remain unanswered. First, is the
distribution of DM on galactic (i.e. kpc) scales really universal?
For example,~\cite{Aden2009} have noted the existence of considerable
scatter in the estimated masses of the lowest luminosity systems, and
several authors have presented evidence of systematic differences
between the properties of the Milky Way dSph satellites and those
surrounding M31~\citep{Collins11,Collins10,Kalirai10}.  Secondly, why
do the properties of the dark and luminous mass distributions appear
to be related, even though baryons dominate, at most, only the inner
regions of galaxies?

The study of the internal kinematics of the Milky Way dSphs has been
revolutionized by the availability of multi-object spectrographs on 4m
and 8m-class telescopes.  Large data sets comprising between several
hundred and several thousand individual stellar velocities per galaxy
have now been acquired for all the luminous dSphs surrounding the
Milky Way \citep{Wilkinson04, Kleyna04, Munoz05,
  Munoz06,walker09a,Koch07a,Koch07b,Battaglia08}.  The volume of the
currently available data is sufficient to measure the dynamical masses
interior to the stellar distributions of the dSphs.  However, the mass
profiles remain less well-determined.  It has been demonstrated
\citep{Walker07,Koch07a,Koch07b,Battaglia08} that these profiles are
consistent with the cuspy dark matter haloes produced in cosmological
N-body simulations (e.g., \citealt{navarro97}), as well as with more
general families of haloes that range from centrally cored to steeply
cusped \citep{walker09b}.  The velocity dispersion profiles alone
cannot distinguish between cored and cusped haloes due to the
degeneracy between mass and velocity anisotropy~\citep[see
  e.g. ][]{Koch07a, Battaglia08,Evans09}. \cite{Gilmore07} recently
showed that the kinematic data and additional features in a small
number of the well-studied dSphs are consistent with cored DM
potentials, under the assumptions of spherical symmetry and velocity
isotropy. In addition, several authors have presented arguments which
suggest that the internal kinematics of dSphs may be more consistent
with cored
haloes~\citep{kleyna02,Goerdt06,Battaglia08,Amorisco11,Walker11}.

In the present paper, we investigate whether DM haloes of Burkert form
are consistent with the observed kinematics of the luminous Milky Way
dSphs. As noted above, Burkert haloes provide good matches to the
rotation curves of spiral galaxies and it is therefore interesting to
ask whether they are also relevant models for galaxies of other Hubble
Types. We also wish to explore whether the parameters of the best-fit
Burkert profiles for the dSphs lie on the extrapolation to the dSph
regime of the scaling relations seen in spiral galaxies. While
previous work has already shown that the internal kinematics of the
Milky Way dSphs may be consistent with cored haloes \citep{Gilmore07},
it does not necessarily follow that Burkert profiles in particular
reproduce the observed kinematic data.

In what follows, we assume that the form of the dSph dark matter halo
density is known and only the length scale and density scale can vary.
We allow the velocity anisotropy of the stellar distribution to vary
in order to reproduce the observed dispersion profiles as closely as
possible. Finally, we compare the resulting DM structural parameters
with the low-luminosity extrapolation of the relations between the
equivalent parameters found in spirals.

Some of the comparisons between dSphs and spirals require us to define
a stellar length scale for the dSphs which plays the same role as the
disk scale length $R_{\rm D}$ in spirals.  One way to do this is to
identify the location of the peak in the circular velocity curves in
each system which would be predicted if the stellar components were
assumed to contribute all the gravitating mass. For a Freeman disk,
this peak occurs at $2.2\ R_{\rm D}$. If the stellar components of the
dSphs are modelled using a~\cite{Plummer1915} sphere the corresponding
radius is at 1.4 $r_{\rm h}$ ($r_{\rm h}$ is the projected half-light
radius). Thus, where necessary, we associate the spiral disk length
scale $R_{\rm D}$ with the radius 0.64 $r_{\rm h}$ in the dSphs.
However, we note that most of our conclusions in this paper do not
make use of this length scale.

The outline of the paper is as follows. In Section~\ref{sec:data}, we
summarize the observational data used in our study and describe in
detail the analysis of the dSph data.  Section~\ref{sec:dmprop}
compares the properties of the dark haloes of spiral and dSph
galaxies. Section~\ref{sec:conc} summarizes our findings and
speculates on their implications for our understanding of dark matter
and galaxy formation.

\section{Data}
\label{sec:data}

\subsection{dSph galaxies}

\begin{figure*}
\begin{center}
\includegraphics[height=14.0truecm]{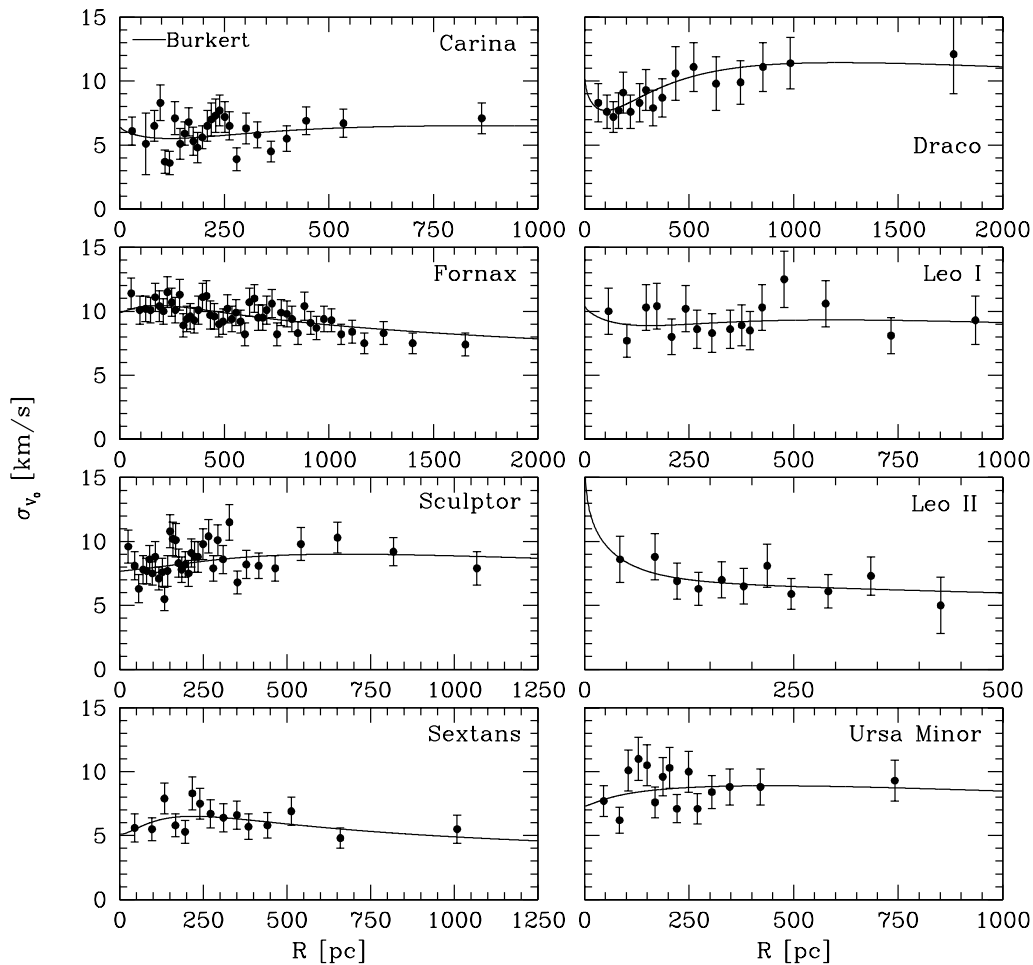}
\end{center}
\caption{Velocity dispersion profiles for the Milky Way's eight
  ``classical'' dSph satellites.  Over-plotted are the best-fitting
  profiles obtained under the assumptions of Burkert dark matter
  haloes, Plummer light profiles and radially constant velocity
  anisotropy. The parameters of each fit, together with associated
  confidence limits, are listed in Table \ref{tab:mcmcresults}.}
\label{fig:burkert_profiles}
\end{figure*}

Figure \ref{fig:burkert_profiles} displays empirical velocity
dispersion profiles originally published by
\cite{Walker07,Mateo08,walker09a} and \cite{walker09b} for the Milky
Way's ``classical'' dSph satellites Carina, Draco, Fornax, Leo I, Leo
II, Sculptor, Sextans and Ursa Minor. As discussed above, in our
present analysis, we assume Burkert profiles for the dark matter
haloes of the dSphs in order to provide a basis for comparison with
spiral galaxies.  Specifically, we use the previously published
velocity dispersion profiles shown in Figure
\ref{fig:burkert_profiles} to constrain values of the Burkert
parameters $\rho_0$ and $r_0$ for each dSph in our sample.

We assume that the luminous component of each dSph consists of a
single, pressure-supported stellar population that is in dynamical
equilibrium and therefore traces the underlying gravitational
potential which we assume to be dominated by the dark matter halo. The
masses of these stellar spheroids can be estimated from their
luminosities: they are 1-2 orders of magnitude smaller than the
dynamical masses. We assume that the stellar mass-to-light ratios
(M/L)$_{\rm V}$ are unity. While the actual (M/L)$_{\rm V}$ ratios may
vary by about a factor of two depending on the details of the stellar
populations~\citep[see e.g.][]{Mateoetal98}, the uncertainties in the
modelling results are dominated by the unknown velocity
anisotropy. 

The Jeans equation relates the density and velocity dispersion of the
stellar component to the mass profile of the dark matter halo.
Assuming spherical symmetry, the Jeans equation for a non-rotating
system can be written \citep{bt08}
\begin{equation}
  \frac{1}{\nu}\frac{d}{dr}(\nu \bar{v_r^2})+2\frac{\beta\bar{v_r^2}}{r}=-\frac{GM(r)}{r^2},
  \label{eq:jeans}
\end{equation}
where $\nu(r)$, $\bar{v_r^2}(r)$, and $\beta(r)\equiv
1-\bar{v_{\theta}^2}/\bar{v_r^2}$ represent the 3-dimensional density,
radial velocity dispersion, and orbital anisotropy, respectively, of
the stellar component, and $M(r)$ is the mass profile of the dark
matter halo.  In this analysis the orbital anisotropy $\beta(r)$ is
not constrained, as all information about the velocity distribution is
restricted to the component along the line of sight.  We make the
simplifying assumption that $\beta=\mathrm{constant}$, which provides
the following solution to Equation \ref{eq:jeans} \citep{mamon05}:
\begin{equation}
  \nu\bar{v^2_r}=Gr^{-2\beta}\displaystyle\int_r^{\infty}s^{2\beta-2}\nu(s)M(s)ds.
  \label{eq:jeanssolution}
\end{equation}
In order to compare to observables, we consider the projection of
Equation \ref{eq:jeanssolution} along the line of sight \citep{bt08}:
\begin{equation}
  \sigma_p^2(R)=\frac{2}{I(R)}\displaystyle \int_{R}^{\infty}\biggl (1-\beta\frac{R^2}{r^2}\biggr ) \frac{\nu \bar{v_r^2}r}{\sqrt{r^2-R^2}}dr,
  \label{eq:jeansproject}
\end{equation}
where $I(R)$ is the projected stellar density profile and
$\sigma_p(R)$ is the projected velocity dispersion profile.  The two
parameters of interest are of course the central density and core
radius, which enter Equation \ref{eq:jeansproject} upon substituting
for $\nu\bar{v_r^2}$ (Equation \ref{eq:jeanssolution}) with the mass
profile derived from the Burkert density profile:
\begin{eqnarray}
  M(r)&=&4\pi\displaystyle\int_{0}^{r}s^2\rho(s)ds\nonumber\\
 & =&\pi\rho_0r_0^3\left(\ln[(1+r/r_0)^2(1+r^2/r_0^2)]-2\tan^{-1}[r/r_0]\right).
  \label{eq:jeansmass}
\end{eqnarray}

To describe the stellar density profile we adopt a Plummer profile,
$I(R)=L(\pi r_{\rm h}^2)^{-1}[1+R^2/r_{\rm h}^2]^{-2}$, where $L$ is
the total luminosity and $r_{\rm h}$ is the projected half-light
radius (i.e., the radius of the circle that encloses half of the total
luminosity in projection).  Under the assumption of spherical
symmetry, the corresponding 3-D stellar density is then
$\nu(r)=3L(4\pi r_{\rm h}^3)^{-1}[1+r^2/r_{\rm h}^2]^{-5/2}$.
Following \citet{walker09b}, for the eight dSphs considered here we
adopt the V-band luminosities and half-light radii from \citet{IH95}.
All values are tabulated in Table 1 of \citet{walker10}.

Treating the stellar density as a known function, we fit halo models
to the empirical velocity dispersion profiles using the set of three
free parameters: $\vec{\theta}\equiv \{\theta_1,\theta_2,\theta_3\}=\{
\log_{10}[r_0/\mathrm{pc}],
\log_{10}[\rho_0/(\mathrm{M_{\odot}pc^{-3}})],-\log_{10}(1-\beta)\}$.
We adopt uniform priors over the ranges $-10\leq
\log_{10}[\rho_0/(\mathrm{M_{\odot}pc^{-3}})]\leq 5$, $-2\leq
\log_{10}[r_0/\mathrm{pc}]\leq 5$, and $-1\leq -\log_{10}(1-\beta)\leq
1$.  For a given point in parameter space, Equation
\ref{eq:jeansproject} specifies the projected velocity dispersion
profile $\sigma_p(R)$.  We compare model profiles to the empirical
velocity dispersion profiles, $\sigma_{V_0}(R)$ (Figure
\ref{fig:burkert_profiles}), by evaluating the likelihood
\begin{equation}
  L(\vec{\theta})=\displaystyle \prod_{i=1}^N \frac{1}{\sqrt{2\pi\mathrm{Var}[\sigma_{V_0}^2 (R_i)]}}\exp\biggl [-\frac{1}{2}\frac{(\sigma_{V_0}^2(R_i)-\sigma_p^2(R_i))^2}{\mathrm{Var}[\sigma_{V_0}^2 (R_i)]}\biggr ],
  \label{eq:likelihood}
\end{equation}
where $\mathrm{Var}[\sigma_{V_0}^2 (R_i)]$ is the square of the error
associated with the square of the empirical dispersion and $N$ is the
number of bins in the dispersion profile.  We obtain (marginalised)
1-D posterior probability distribution functions for each free
parameter using a Markov-Chain Monte Carlo (MCMC) algorithm.
Specifically, we use the same Metropolis-Hastings algorithm
\citep{metropolis53,hastings70} described in detail by
\citet{walker09b}.  In order to account for the error associated with
observational uncertainty in the half-light radius, for each point
sampled in our MCMC chains we scatter the adopted value of $r_{\rm h}$
by a random deviate drawn from a Gaussian distribution with standard
deviation equal to the published error \citep{IH95}.  The probability
distribution functions for the free parameters are thus effectively
marginalised over the range of half-light radii consistent with
observations.

The velocity dispersion profiles corresponding to the
highest-likelihood (Equation \ref{eq:likelihood}) point from each of
our MCMC chains are over-plotted on the empirical profiles in Figure
\ref{fig:burkert_profiles}. These `best fits' demonstrate that Burkert
profiles can provide an excellent description of dSph velocity
dispersion profiles.  For each free parameter, we take the 1-D
posterior probability distribution obtained from our MCMC chains as
the observational constraint, given our modelling assumptions.  For
each free parameter (and combinations thereof), Table
\ref{tab:mcmcresults} identifies the median value and confidence
intervals that enclose the central 68\% (and 95\%) of accepted points
in our chains. We find that the dSph haloes have central densities
ranging from $7\times 10^{-24}$g cm$^{-3}$ to $3\times 10^{-22}$g
cm$^{-3}$ and core radii ranging from $0.05$ kpc to $0.65$ kpc. The
data in the table also exhibit the well-known mass-anisotropy
degeneracy: because the dispersion profiles of the dSphs are
essentially flat to large radii and have a relatively small range of
amplitudes, larger values of $\rho_0r_0^3$ are required for galaxies
with more radially biased velocity distributions. Our analysis is
particularly susceptible to this degeneracy due to our restriction of
the modelling to Burkert halo profiles.

We have repeated our analysis with the additional assumption of
velocity isotropy (i.e. $\beta=0$). With the exception of Sextans, the
halo parameters obtained for our sample are consistent with those in
Table~\ref{tab:mcmcresults} within the quoted uncertainties. When we
restrict ourselves to isotropic models, the best-fit $r_0$ for Sextans
falls to the significantly smaller value of $47$pc. This is consistent
with the fact that Sextans is unique in requiring tangential
anisotropy to obtain a good fit to the dispersion profile - the
best-fit isotropic model does not match the profile of Sextans
interior to $200$pc.

\begin{table*}
\begin{center}
\scriptsize
\begin{tabular}{lclrrrrrrr}
\\\hline\hline
Object &L$_V$/L$_{V,\odot}$&r$_{\mathrm{h}}$/pc&$-\log_{10}[1-\beta]$&$\log_{10}[\rho_0/(\mathrm{M_{\odot}pc^{-3}})]$&$\log_{10}[r_0/\mathrm{pc}]$&$\log_{10}[\rho_0 r_0/\mathrm{(M_{\odot }pc^{-2})}] $&$\log_{10}[\mathrm{M_T(R_{83}/2)/M_{\odot}}]$&$\log_{10}[\mathrm{M_s(R_{83}/2)/M_h(R_{83}/2)}]$ \\\hline
Carina    &(2.4 $ \pm $ 1.0)$\ \times$10$^5$&241$\pm$23&$ 0.18_{-0.26(-0.60)}^{+ 0.22(+ 0.47)}$&$-1.19_{-0.22(-0.43)}^{+ 0.31(+ 0.75)}$&$ 2.78_{-0.28(-0.59)}^{+ 0.28(+ 0.64)}$&$ 1.62_{-0.05(-0.09)}^{+ 0.07(+ 0.22)}$&$ 7.07_{-0.12(-0.26)}^{+ 0.08(+ 0.12)}$&$-1.71_{-0.08(-0.12)}^{+ 0.12(+ 0.27)}$\\ \\
Draco     &(2.7 $ \pm $ 0.4)$\ \times$10$^5$&196$\pm$12&$0.26_{-0.29(-0.76)}^{+ 0.35(+ 0.66)}$&$-0.74_{-0.17(-0.30)}^{+ 0.22(+ 0.56)}$&$ 2.81_{-0.23(-0.51)}^{+ 0.21(+ 0.40)}$&$ 2.09_{-0.05(-0.10)}^{+ 0.06(+ 0.14)}$&$ 7.31_{-0.11(-0.20)}^{+ 0.09(+ 0.17)}$&$-2.17_{-0.09(-0.17)}^{+ 0.11(+ 0.20)}$\\ \\
Fornax   &(1.4 $ \pm $ 0.4)$\ \times$10$^7$&668$\pm$34&$-0.07_{-0.09(-0.24)}^{+ 0.07(+ 0.14)}$&$-0.72_{-0.16(-0.30)}^{+ 0.21(+ 0.57)}$&$ 2.57_{-0.12(-0.30)}^{+ 0.09(+ 0.18)}$&$ 1.85_{-0.07(-0.12)}^{+ 0.09(+ 0.27)}$&$ 8.10_{-0.02(-0.04)}^{+ 0.02(+ 0.03)}$&$-1.16_{-0.02(-0.03)}^{+ 0.02(+ 0.04)}$\\ \\
Leo I      &(3.4 $ \pm $ 1.1)$\ \times$10$^6$&246$\pm$19&$-0.00_{-0.47(-0.92)}^{+ 0.37(+ 0.79)}$&$-0.39_{-0.35(-0.55)}^{+ 0.52(+ 1.30)}$&$ 2.45_{-0.33(-0.73)}^{+ 0.28(+ 0.51)}$&$ 2.07_{-0.08(-0.13)}^{+ 0.18(+ 0.56)}$&$ 7.58_{-0.07(-0.18)}^{+ 0.05(+ 0.10)}$&$-1.15_{-0.06(-0.10)}^{+ 0.08(+ 0.19)}$\\ \\
Leo II     &(5.9 $ \pm $ 1.8)$\ \times$10$^5$&151$\pm$17&$-0.15_{-0.59(-0.81)}^{+ 0.71(+ 1.06)}$&$ 0.61_{-0.92(-1.33)}^{+ 2.12(+ 3.80)}$&$ 1.76_{-0.92(-1.54)}^{+ 0.59(+ 0.96)}$&$ 2.39_{-0.33(-0.47)}^{+ 1.18(+ 2.25)}$&$ 7.15_{-0.09(-0.21)}^{+ 0.09(+ 0.18)}$&$-1.65_{-0.10(-0.18)}^{+ 0.09(+ 0.21)}$\\ \\
Sculptor &(1.4 $ \pm $ 0.6)$\ \times$10$^6$&260$\pm$39&$ 0.01_{-0.12(-0.25)}^{+ 0.11(+ 0.19)}$&$-0.61_{-0.15(-0.27)}^{+ 0.16(+ 0.34)}$&$ 2.55_{-0.11(-0.23)}^{+ 0.13(+ 0.22)}$&$ 1.96_{-0.04(-0.07)}^{+ 0.04(+ 0.10)}$&$ 7.55_{-0.04(-0.10)}^{+ 0.03(+ 0.06)}$&$-1.48_{-0.03(-0.06)}^{+ 0.04(+ 0.10)}$\\ \\
Sextans  &(4.1$ \pm $ 1.9)$\ \times$10$^5$&682$\pm$117&$-0.64_{-0.25(-0.35)}^{+ 0.37(+ 0.60)}$&$ 0.74_{-1.24(-1.97)}^{+ 1.39(+ 3.02)}$&$ 1.67_{-0.55(-1.17)}^{+ 0.54(+ 0.93)}$&$ 2.41_{-0.70(-1.03)}^{+ 0.85(+ 1.87)}$&$ 7.51_{-0.09(-0.17)}^{+ 0.09(+ 0.18)}$&$-2.09_{-0.10(-0.18)}^{+ 0.09(+ 0.17)}$\\ \\
Ursa Minor&(2.0 $ \pm $ 0.9)$\ \times$10$^5$&280$\pm$15&$-0.20_{-0.47(-0.74)}^{+ 0.25(+ 0.50)}$&$-0.39_{-0.45(-0.79)}^{+ 0.75(+ 1.39)}$&$ 2.39_{-0.44(-0.74)}^{+ 0.36(+ 0.77)}$&$ 2.00_{-0.09(-0.15)}^{+ 0.31(+ 0.63)}$&$ 7.56_{-0.07(-0.20)}^{+ 0.05(+ 0.09)}$ &$-2.38_{-0.05(-0.09)}^{+ 0.07(+ 0.20)}$\\\hline
\end{tabular}
\caption{Table of dSph structural parameters and results of our mass
  modelling. The columns are: (1-3) dSph name, observed V-band
  luminosity and half-light radius~\protect\citep[from ][]{IH95}; (4)
  velocity anisotropy; (5-6) density scale and length scale for the DM
  halo, assuming a Burkert DM density profile; (7) halo ``surface
  density'' scale; (8) total mass interior to $R_{83}$/2, where
  $R_{83}$ is the radius enclosing 0.83 of the total luminosity; (9)
  ratio of stellar mass to DM mass interior to $R_{83}$/2. In columns
  (4-9) the median value of the parameter is given together with error
  bars which enclose the central $68\%$ ($95\%$) of the marginalised
  1D probability distribution function for the parameter. Since we
  assume that M/L$_V$=1 for the stellar components of the dSphs, the
  luminosities directly yield the stellar masses.}
\label{tab:mcmcresults}
\end{center}
\end{table*}

 \begin{figure}
\begin{center}
\vskip -2.8truecm
\psfig{file=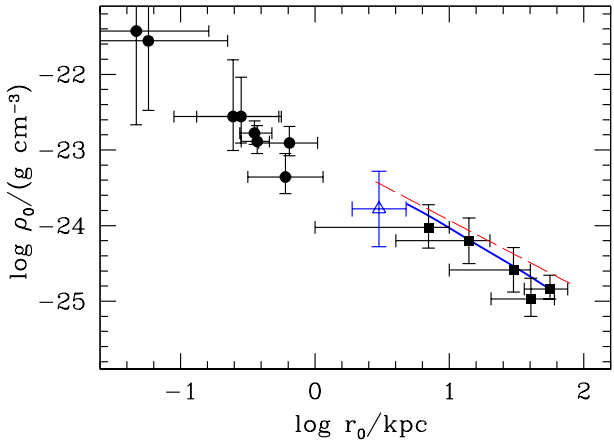,width=8.8truecm}
\end{center}
\vskip -0.3truecm
\caption{Parameters of Burkert DM haloes obtained from dynamical
  modelling of (i) spirals, based on the URC hypothesis applied to
  co-added rotation curves (solid line; data from PSS) or weak lensing
  shear~\protect\citep[squares; ][]{Hoekstra05}; (ii) NGC 3741
  (triangle) the darkest spiral in the local Universe based on its
  kinematics; (iii) the ``classical'' Milky Way dSph satellites
  (filled circles), based on their internal stellar kinematics (this
  paper). The~\protect\cite{spano} relation is shown as a dashed
  line.}
\label{fig:rho0r0}
\end{figure}

\subsection{Spiral Galaxies}

As discussed above, Burkert halo models provide excellent fits to
individual spiral galaxy rotation curves as well as to samples of
co-added rotation curves. Moreover, when the mass modelling is
performed using Burkert haloes, a tight relation between $\rho_0$ and
$r_0$ (and also between other parameters like the disk and virial
masses) emerges~\citep[PSS; ][]{Donato04,Salucci07}.  As can be seen
in Figure~\ref{fig:rho0r0}, we find similar $\rho_0$ vs $r_0$
relationships {\em independently} of whether the mass profiles are
obtained from rotation curves or from gravitational lensing data and
irrespective of whether the analysis is performed on individual or
co-added rotation curves.
 
\begin{figure}
\begin{center}
\vskip -2.8truecm
\psfig{file=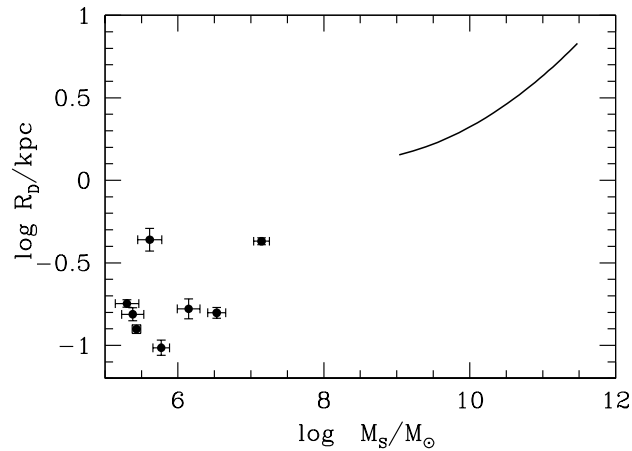,width=8.8truecm  }
\end{center}
\vskip -0.3truecm
\caption{Comparison of the distribution of characteristic baryonic
  scale $R_{\rm D}$ versus stellar mass $M_{\rm s}$ for dSphs (points;
  this paper) with the corresponding relation in Spirals (from
  PSS). See Section~\ref{sec:intro} for the definition of $R_{\rm D}$
  used for the dSph sample.}
\label{fig:r83}
\end{figure}

To emphasise the very different ranges of baryonic mass and extent
probed by the dSphs and the spiral galaxies in our sample, in
Figure~\ref{fig:r83}, we compare the relationship between the
characteristic baryonic length scale ($R_{\rm D}$; see above for
definitions) and the stellar mass of dSphs (estimated from the $V$
band luminosity, assuming a stellar mass-to-light ratio of unity [in
solar units]), and of spirals. 

\section{Dark matter scaling relations}
\label{sec:dmprop}
In the previous section, we showed that Burkert halo profiles provide
good fits to the dSph kinematic data
(Figure~\ref{fig:burkert_profiles}). In this section we compare the
parameters of the Burkert profiles obtained from our dynamical
modelling of dSphs with those obtained for spiral galaxies.  In
Figure~\ref{fig:rho0r0}, we plot $\rho_0$ versus $r_0$ for the eight
dSphs in our sample. Remarkably, they lie on the extrapolation to
higher central densities of the relation found for spirals.  All these
data can be reproduced by the relation $\log\: \rho_0 \simeq \alpha
\:\log\: r_0$ +const with $0.9 <\alpha<1.1$.

\begin{figure}
\begin{center}
\vskip -3.7truecm
\psfig{file=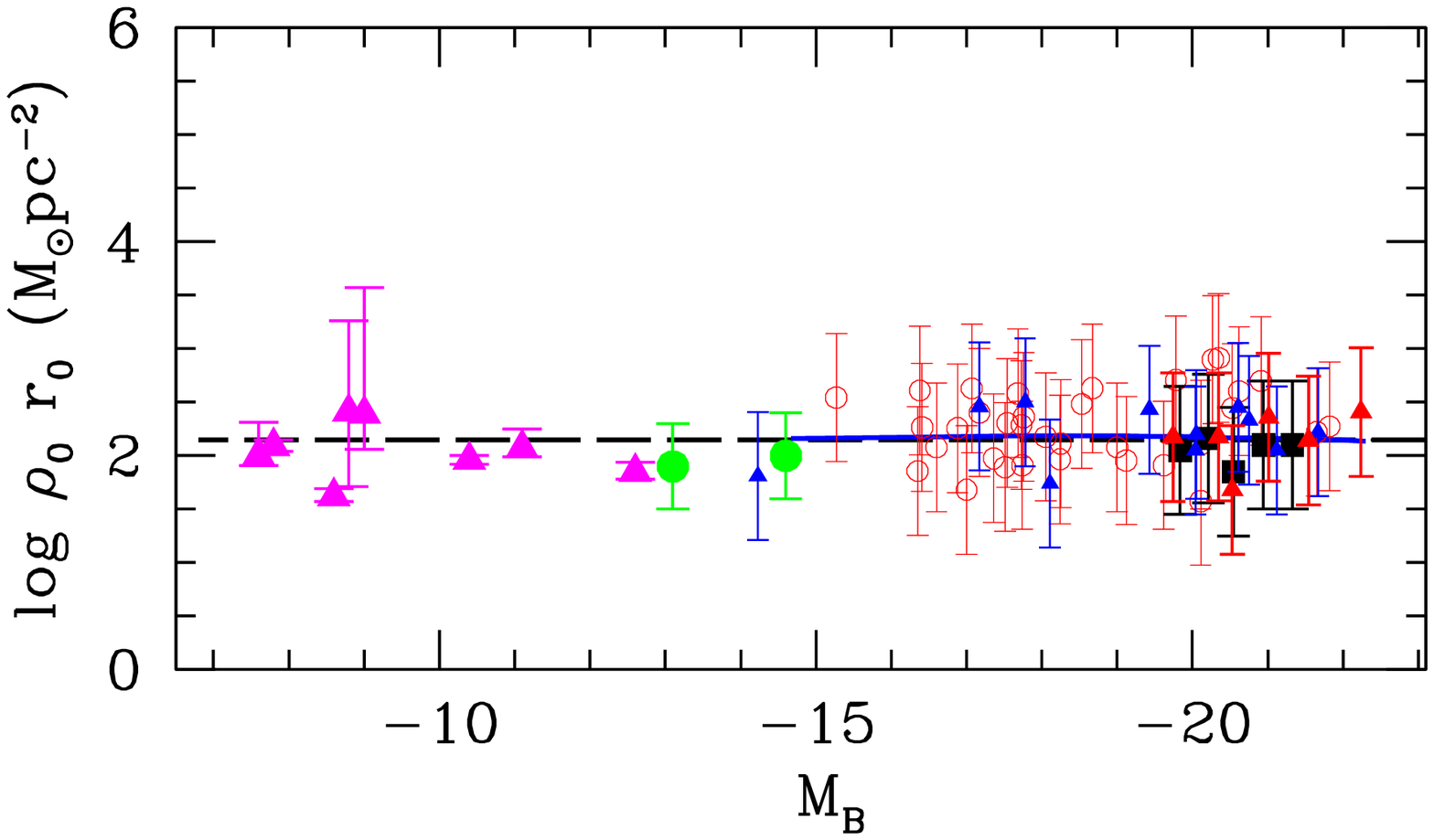,width=9truecm}
\end{center}
\vskip -0.8truecm
\caption{$\rho_0 r_0$ in units of M$_\odot$ pc$^{−2}$ as a function of
  galaxy magnitude for different galaxies and Hubble types. The data
  are: (1) the~\protect\cite{spano} sample of spiral galaxy data (open
  red circles); (2) the URC relation~\citep[solid blue
  line;][]{shankar}; (3) the dwarf irregulars N
  3741~\protect\citep[M$_{\rm B}$ = −13.1; ][]{g07} and DDO
  47~\protect\citep[M$_{\rm B}$ = −14.6; ][]{g05} (full green
  circles), spirals and ellipticals~\protect\citep[black squares;
  ][]{Hoekstra05} investigated by weak lensing; (4) Milky Way dSphs
  (pink triangles - this paper); (5) nearby spirals in THINGS~\protect\citep[small
  blue triangles; ][]{Walter08}; (6) early-type
  spirals~\protect\citep[full red triangles; ][]{Noor06,Noor07}.  The
  long-dashed line shows the~\protect\cite{donato09} result.}
\label{fig:sigmacost}
\end{figure}

An even more interesting comparison can be done involving the mean
dark matter surface density within the dark halo core radius (the
radius within which the volume density profile of dark matter remains
approximately flat).  It was recently discovered~\citep{donato09} that
this quantity $\mu_0$ ($\mu_0 \equiv\rho_0 r_0$) is constant across a
wide range of galaxies of different Hubble Type and luminosity and
that this relation holds also for dSphs, as we confirm in
Figure~\ref{fig:sigmacost} where the data for the dSphs are the
results of our Burkert halo modelling in the present paper (in
\citeauthor{donato09} the dSph halo parameters were obtained via a
different approach).  We therefore confirm that this relation extends
across a luminosity range of 14 magnitudes and spans the whole Hubble
sequence.  The potential implications of the constancy of $\mu_0$ are
discussed in~\cite{donato09}.

\begin{figure}
\begin{center}
\vskip -2.8truecm
\psfig{file=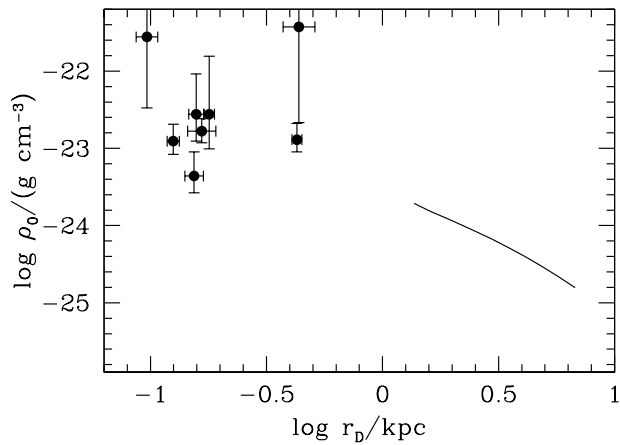,width=8.8truecm}
\end{center}
\vskip -0.3truecm
\caption{Halo central density $\rho_0$  versus stellar length  
scale $R_{\rm D}$ for spirals (solid curve) and dSphs (points).}
\vskip -0.0truecm
\label{fig:rho0RD}
\end{figure}

In their modelling of spiral galaxies using the URC hypothesis, PSS
found that the parameters of their Burkert DM haloes were correlated
with those of the luminous matter. In Figure~\ref{fig:rho0RD} we show
the $\rho_0$ vs $R_D$ relationship for our dSph sample compared with
the corresponding relation for spirals from PSS. As in the case of the
$\rho_0$-$r_0$ relation in Figure~\ref{fig:rho0r0}, the dSph data are
consistent with the extrapolation of the relation seen in spirals.
The significance of this relation derives from the fact that it links
the DM and baryonic matter properties of galaxies on a wide range of
length scales: qualitatively, the ``central'' densities of DM haloes
increase as the extents of their associated stellar components decrease
Although the observational evidence for this relation is relatively
strong, we stress that its physical interpretation is presently
unknown~\citep[see ][for some related discussion of this point]{Gentile2009,Angus2008}.

It is interesting to consider explicitly the role played by velocity
anisotropy in this result. Unsurprisingly, under the assumption of
Burkert haloes for dSphs the inclusion of velocity anisotropy as a
free parameter improves the quality of the dispersion profile fits
relative to those obtained for isotropic models. However, we also find
that the scatter in the $\rho_0$-$r_0$ relation is smaller for
anisotropic models -- thus the better we reproduce the observed
dispersion profiles using Burkert haloes, the tighter is the
correlation between the halo parameters. 

Finally, we emphasise that the present results, do not {\it require}
the presence of cored haloes in dSphs, nor do they constrain the
density and scale lengths of their haloes in a model-independent way.
On the other hand, the fact that the dSph kinematics can be reproduced
using Burkert DM halo profiles whose structural parameters lie on the
same scaling relations as those of spirals provides some support for
the claim that the mass distributions in dSph galaxies can be
understood within the same framework as those of spirals.

\section{Conclusions}
\label{sec:conc}
Dwarf spheroidal galaxies are the lowest luminosity stellar systems
which show evidence of dynamically significant DM. Their physical
properties (luminosity, stellar scale length, baryon fraction) are
typically two orders of magnitude different from those observed for
spiral and elliptical galaxies. Given these extreme structural
properties, an understanding of the formation of dSphs is crucial for
the development of a complete picture of galaxy formation.

The main result of this paper is the finding that these galaxies,
despite being very distinct in their physical properties from spirals
and ellipticals and having a large individual scatter in their
baryonic properties, exhibit kinematic properties that can be modelled
using DM haloes with the same mass profiles as those which reproduce
the rotation curves of spiral galaxies. Under the assumption that the
haloes of dSphs have Burkert profiles, we find that the derived
central densities and the core radii are consistent with the
extrapolation of the relationship between these quantities seen in
spiral galaxies.  Conversely a Burkert profile with structural
parameters predicted by the extrapolation of the relation between halo
central density and DM core radius previously found from Burkert fits
to the kinematics of elliptical and spiral galaxies can account for
the observed internal kinematics in dSphs.

This result is intriguing, and could point to a common physical
process responsible for the formation of cores in galactic haloes of
all sizes, or to a strong coupling between the DM and luminous matter
in dSphs.  It is worth noting that a potential connection between
spiral galaxies and dSphs does not appear as natural as one between
dSphs and other hot, spheroidal systems
\citep{Dabringhausen2008,Forbes2008}.  For example, while the sizes of
spiral galaxies are presumably fixed by the angular momentum of the
gas from which they form, most of the present-day dSphs show no signs
of rotation (although \citet{Battaglia08} have recently found evidence
of rotation in the Sculptor dSph).  However, \cite{Mayer2001} have
proposed a formation scenario for dSphs in which they are initially
low-mass disk galaxies that are subsequently transformed into
spheroids by tidal interaction with the Milky Way. More recently, such
models have been shown to provide reasonable models for the properties
of the Fornax \citep{Klimentowski2007} and LeoI \citep{Lokas2008}
dSphs.  If the haloes of dSphs do indeed follow the scaling laws
defined by more massive disk galaxies, this could lend indirect
support to evolutionary histories of this kind. Suggestive evidence of
such transformation scenarios is also provided by the discovery of
residual disks with spiral structure in luminous dwarf elliptical
galaxies in the Virgo cluster~\citep{Lisker2006}.

Further dynamical analysis is needed to derive the actual DM
distribution in dSph and possibly to estimate their halo core radii.
Nevertheless, it interesting to speculate on the possible implications
of these scaling laws for our understanding of DM.  Warm dark matter
has been invoked as a potential solution to the over-prediction of
substructure by $\Lambda$CDM simulations, and to the cusp-core
issue~\cite[e.g.][]{Moore1999}.  However, the existence of scaling
relations between the central density and core radius over three
orders of magnitude in both quantities would argue against this
explanation, unless the warm DM spectrum is extremely fine-tuned.
Further, such DM relations cannot arise due to either
self-annihilation or decay of DM which would predict a narrow range in
$\rho_0$ and no clear correlation of the latter with the core radius.

\cite{Dalcanton2001} argued that the phase-space densities of DM
haloes suggested that warm DM (either collisional or collisionless)
could not be the cause of cores in galaxy haloes on all scales.  These
authors suggested a dynamical origin for the cores of larger galaxies.
Subsequently, a number of studies have demonstrated that macroscopic
core formation in galaxy haloes is possible through the infall of
compact baryonic (or baryon dominated)
sub-clumps~\citep{ElZant2001,Jardel2009,Goerdt2010,Cole2011}. Further
work is required to explore whether such processes, in conjunction
with subsequent star formation and
feedback~\citep[e.g.][]{Pasetto2010}, can result in universal scaling
relations spanning three orders of magnitude in density and length
scales.

Clearly, a direct kinematic determination of the dark matter profiles
of dSphs is essential to confirm the robustness of the scaling
relations between the halo parameters. A number of recent papers have
made progress towards this goal in the subset of dSphs which exhibit
kinematically distinct sub-populations in their stellar
components~\citep{Battaglia08,Amorisco11,Walker11}. Interestingly, all
three analyses favour cored haloes over cusped ones. However, further
observational and modelling work is required to constrain the halo
profiles more tightly.  Theoretically, the development of a physical
picture of the processes which shape the halo profiles of dSphs and
which could lead to the existence of apparently similar scaling
relations between halo properties over a wide range of galaxy
luminosities, is an important research goal for the coming years.

\section*{Acknowledgments}
MIW acknowledges support from a Royal Society University Research
Fellowship. MGW is supported by NASA through Hubble Fellowship grant
HST-HF-51283.01-A, awarded by the Space Telescope Science Institute,
which is operated by the Association of Universities for Research in
Astronomy, Inc., for NASA, under contract NAS5-26555. EKG is supported
by Sonderforschungsbereich SFB 881 ``The Milky Way System''
(subproject A2) of the German Research Foundation (DFG)." AK thanks
the Deutsche Forschungsgemeinschaft for funding from Emmy-Noether
grant Ko 4161/1.

{}

\end{document}